\documentclass[a4paper,12pt]{article}
\usepackage{amsmath}
\usepackage{authblk}
\usepackage{graphicx,epsfig}

\usepackage{color}

\numberwithin{equation}{section}

\title{Gravitational memory for uniformly accelerated observers}

\author{Sanved Kolekar $^{1,2}$ \footnote{sanved.kolekar@nottingham.ac.uk} \; }
\author{\; Jorma Louko$^{1}$ \footnote{jorma.louko@nottingham.ac.uk}}

\affil{$^{1}$ School of Mathematical Sciences, 
University of Nottingham,\\ 
Nottingham NG7 2RD, 
UK}

\affil{$^{2}$ UM-DAE Centre for Excellence in Basic Sciences, \\  Mumbai 400098, India}

\date{March 2017; revised July 2017}
\begin{document}
\maketitle
\begin{abstract}
Recently, Hawking, Perry and Strominger described a physical process 
that implants supertranslational hair on a Schwarzschild black hole 
by an infalling matter shock wave without spherical symmetry. 
Using the BMS-type symmetries of the Rindler horizon, 
we present an analogous process that implants 
supertranslational hair on a Rindler horizon by a matter shock wave 
without planar symmetry, 
and we investigate the corresponding 
memory effect on the Rindler family of uniformly linearly accelerated observers. 
We assume each observer to remain linearly uniformly accelerated through the wave, in 
the sense of the curved spacetime generalisation of the 
Letaw-Frenet equations. Starting with a family of observers who follow the orbits of a 
single boost Killing vector before the wave, we find that after the wave has passed, 
each observer still follows the orbit of a boost Killing vector but 
this boost differs from trajectory to trajectory, and the trajectory-dependence carries 
a memory of the planar inhomogeneity of the wave.
We anticipate this classical memory phenomenon to have a counterpart in  
Rindler space quantum field theory.

\end{abstract}
\section{Introduction}

Recently, Hawking, Perry and Strominger \cite{HPS1,HPS2} (HPS) 
have shown that a black hole in an asymptotically flat spacetime has an infinite collection of soft hairs corresponding to the infinite supertranslation symmetries of the flat spacetime at asymptotic infinity. These supertranslations are essentially diffeomorphisms on the spacetime which leave 
the asymptotic structure at null infinity intact and belong to the 
Bondi-Metzner-Sachs (BMS)
subgroup \cite{BMS}. Classically, diffeomorphisms do not affect the vacuum associated with the phase space of the canonically conjugate variables of gravitational degrees of freedom. However, from a field theoretic perspective, it has been argued that the supertranslations act non-trivially on the degenerate vacua related to the infinite BMS asymptotic symmetries and are spontaneously broken, accompanied
by the creation/annihilation of Goldstone bosons, namely, the soft photons and soft gravitons. 
The results due to Christodoulou and Klainerman \cite{Chris} on stability of Minkowski spacetime and asymptotic boundary conditions allow one to construct an infinite number of 
nonvanishing
conserved supertranslation and superrotation charges on 
the past and future null infinities of generic asymptotically flat
spacetimes. For the black hole spacetimes, as considered by HPS, it has been
conjectured \cite{HPS1} that these charges would enable the outgoing Hawking quanta to 
contain enough correlations to make the evaporation unitary.
(Also see~\cite{Mirb}.)

It was shown in \cite{HPS2} that soft hair can be implanted on a 
Schwarzschild black hole by a physical process, 
an infalling matter shock wave that does not have spherical symmetry. 
The metric after the wave is related to the metric 
before the wave by a BMS supertranslation. 
As these supertranslations generate non-trivial time 
translations on the null generators of the event horizon, 
they act like a gravitational memory on the horizon. 
This raises the possibility that these horizon 
supertranslations could be the mechanism that encodes 
correlations in the outgoing Hawking quanta~\cite{Hawconf}.

Hawking's prediction of black hole radiation \cite{Hawking}
relies on the semi-classical framework for 
gravity, wherein only the matter fields 
propagating on the classical background geometry are quantised. 
Within the same framework, however with a more practical approach, it is known that an 
Unruh-DeWitt detector
coupled to the Hartle-Hawking state of the quantum field and positioned at a fixed radius outside the hole responds thermally~\cite{BD,Unruhdet}.
In particular, the response rate of the detector is of the 
Kubo-Martin-Schwinger form \cite{Kubo:1957mj,Martin:1959jp,Haag:1967sg}. 
A uniformly linearly accelerated observer/trajectory 
plays a central role in these semi-classical analyses: the thermal form of the Hawking radiation is a peculiarity associated only with the uniformly linearly accelerated observers. 
A freely falling detector, either radially in-falling or on a elliptical/circular orbit responds quite differently albeit non-thermally \cite{Lee}.

An interesting question one would like to address is how does implanting a black hole with supertranslation hair affect the thermal response of the uniformly linearly
accelerated detector. There are two aspects involved in such an investigation. 
First, it is well known that 
the
passage of gravitational radiation, in-falling or outgoing, results in a change in the mutual proper-separation of geodesic observers at asymptotic infinity, an effect called 
the
gravitational memory effect \cite{Zeld}. 
More recently, 
it has been explicitly shown that the memory effect for geodesic observers is equivalent to that of a diffeomorphism on the Schwarzschild metric belonging to the class of BMS supertranslations at asymptotic infinity \cite{Strom-bms1,Winicour-memory,Ashtekar-memory,Strom-zhiboedov}. One 
can expect the supertranslations to have a distinguishable effect at a classical level on the congruence of uniformly accelerated observers as well. 
The second aspect is 
the evolution of a quantum field when a 
supertranslation diffeomorphism is present.
In this paper we focus on the first aspect, for uniformly 
linearly accelerated observers in the supertranslated Schwarzschild 
black hole and in its Rindler analogue. 
We plan to address the related quantum aspects in a future paper \cite{Sanjorma}.

The flat spacetime analogue of Hawking radiation is the Unruh effect \cite{Davies,Unruh}. 
A uniformly linearly
accelerated observer, moving on an integral curve of a boost Killing vector, perceives the Minkowski vacuum to be thermal with a temperature proportional to the magnitude of its acceleration. In contrast to the black hole case, the mode solutions of the quantum field in the Rindler spacetime are known in terms of well studied special functions. The analytical tractability and conceptual similarity often makes 
Rindler spacetime a pre-exploratory arena 
for studying numerous black hole effects, which we also exploit in this paper. We extend the physical process of implanting a supertranslational hair described by HPS to the case of the Rindler 
horizon, and we investigate the corresponding gravitational memory effects on uniformly accelerated observers. The BMS-type horizon symmetries for the Rindler spacetime have been 
found in \cite{Donn, Eling, Cai} (for a related discussion see~\cite{Hotta}). 
In section \ref{hairsection}, we briefly review a class of such Rindler supertranslations and introduce 
an asymmetric 
matter
shock wave impinging on the Rindler horizon. In section \ref{linearsection}, 
we motivate and propose a covariant way to define 
uniformly linearly
accelerated trajectory in curved 
spacetime. 
In section \ref{memorysection}, we analyse the effect of implanting supertranslational hair on 
uniformly linearly
accelerated motion in the Rindler spacetime. 
Starting with a family of trajectories that follow the orbits of a 
single boost Killing vector before the wave, we find that after the wave has passed, 
each trajectory still follows the orbit of a boost Killing vector but 
this boost differs from trajectory to trajectory, and the trajectory-dependence carries 
a memory of the planar inhomogeneity of the wave.
We further show that the effect of supertranslations on 
uniformly linearly
accelerated observers in the Schwarzschild spacetime is even more drastic with the trajectory falling inside the black hole horizon for an ingoing shock wave or the trajectory ejecting out to spatial infinity for a outgoing shock wave. 
Concluding remarks are collected in section \ref{discsection}.

The Minkowski metric is taken to have the mostly plus signature, and Roman indices run over all spacetime indices.

\section{Implanting supertranslational hair to the Rindler horizon} \label{hairsection}

In \cite{HPS2}, HPS considered a linearised
shock wave without spherical symmetry propagating on a Schwarzschild spacetime. The metric for the complete process of implanting the supertranslational hair is given by
\begin{eqnarray}
ds^2 &=& - \left(1 - \frac{2M}{r} - h(v - v_0)\frac{2\mu}{r}  -  h(v - v_0)\frac{MD^2 C}{r^2}\right) dv^2 + 2dv dr \nonumber \\
&& -  h(v - v_0) D_A \left( 2C - \frac{4MC}{r} + D^2C \right) dv d\Theta^A \nonumber  \\
&& + \left( r^2\gamma_{AB} + h(v - v_0) 2rD_A D_B C -  h(v - v_0)r\gamma_{AB} D^2C \right) d\Theta^A \Theta^B \nonumber \\
\label{schwsupermetric}
\end{eqnarray}
where the co-ordinates $(v,r,\Theta^A)$ are the advanced 
Bondi co-ordinates 
(surfaces of constant $v$ are ingoing family of null hypersurfaces), 
$\gamma_{AB} d\Theta^A \Theta^B$ is the metric on
two-sphere, 
$D^A$ is the covariant derivative on the unit two-sphere, 
$h(v-v_0)$ is the Heaviside step function and the function $C(\Theta)$ characterises the angular profile of the shock wave. 
The metric differs from a Schwarzschild metric of mass $M >0$ by
\begin{equation}
h_{ab} = h(v - v_0) \bigg ( {\cal L}_{\Xi}g_{ab} - \frac{2\mu}{r}\delta^v_a \delta^v_b \bigg)
\label{schwpert}
\end{equation}
where  $\Xi^a = \left[C, -D^2C/2, D^AC/r \right]$ 
is the BMS-type supertranslation vector, preserving to linear order the Bondi gauge conditions  
$g_{rr} = 0 = g_{rA}$ 
and $\det\left( g_{AB}/r^2\right) = g(\Theta)$, and 
satisfying the asymptotic fall-off conditions required to preserve the asymptotic infinity. 
The stress-energy tensor of the shock wave is 
\begin{eqnarray}
T_{vv} &=& \frac{1}{4\pi r^2} \left[ \mu + \frac{D^2 \left(D^2 + 2 \right)C }{4} - \frac{3MD^2C}{2r} \right] \delta(v-v_0) \nonumber \\
T_{vA} &=& -\frac{3MD_A C}{8\pi r^2} \delta(v-v_0)
\label{eq:schw-stressenergy}
\end{eqnarray}

\begin{figure}
\includegraphics[width=\linewidth]{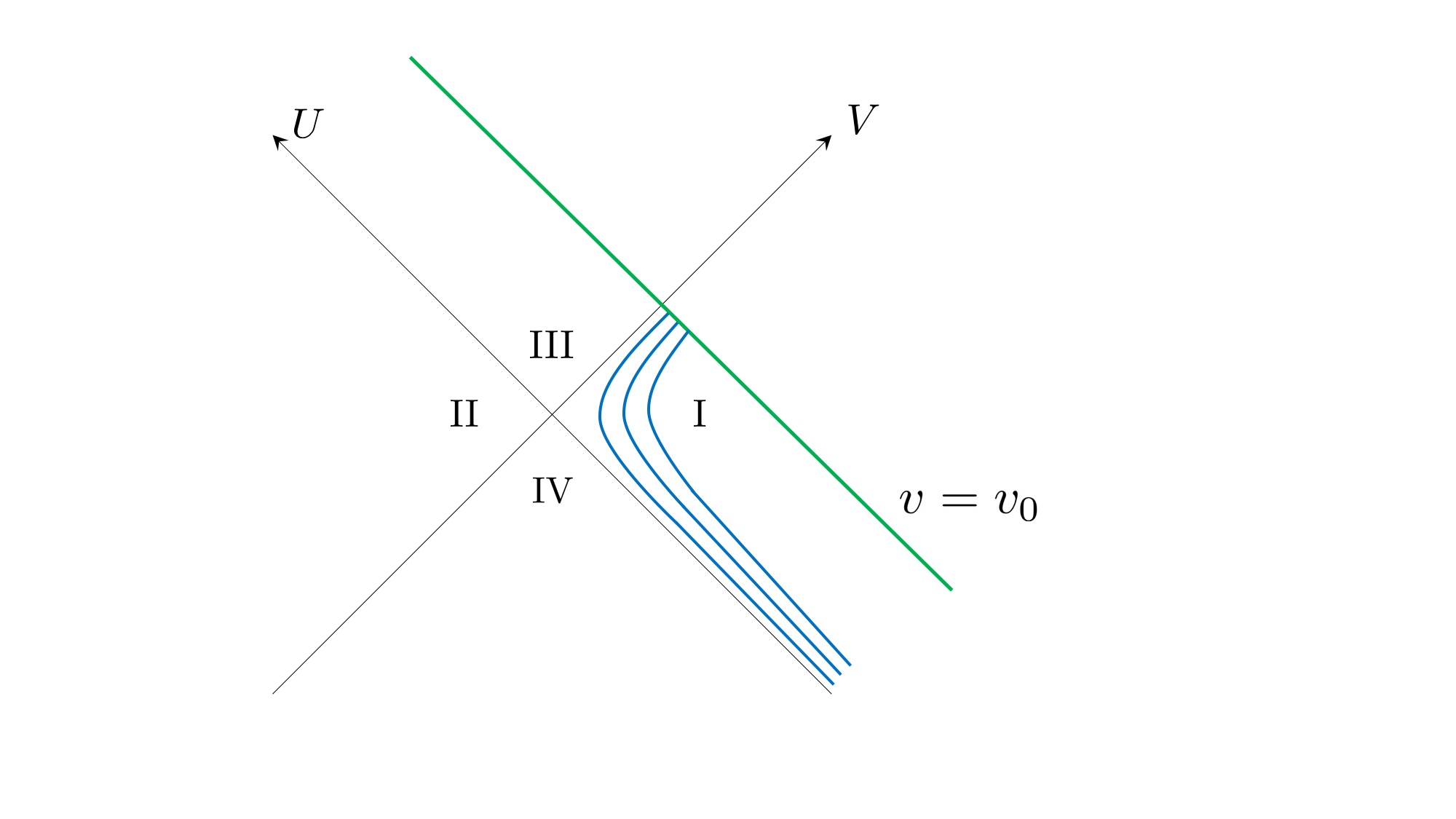}
\caption{The Rindler spacetime. The metric reads 
$ds^2 = - dU\,dV + \delta_{AB}  dx^A dx^B$, and the two 
transverse dimensions $x^A$ are suppressed. 
Regions I and II are the right and left Rindler wedges. 
The co-ordinates \eqref{metricRindler} cover regions I and III 
and their joint boundary, such that $r>0$ in I and $r<0$ in~III. 
Curves of constant $v$ are lines of constant~$V$, 
and curves of constant positive $r$ in I and III are hyperbolas of 
constant~$UV$. 
The infalling null shock wave $v=v_0$ is shown, 
with a selection of hyperbolas of constant positive $r$ at $v<v_0$.
\label{fig:diagram1}}
\end{figure}

Similarly to the Schwarzschild case, we consider a 
physical process version of implanting supertranslation hair to the Rindler horizon. 
We begin by writing the metric of the Rindler spacetime in the 
advanced Bondi-type co-ordinates $(v,r,x,y)$ as
\begin{equation}
ds^2 = - 2\kappa r dv^2 + 2 dvdr + \delta_{AB}  dx^A dx^B 
\label{metricRindler}
\end{equation}
where $\kappa >0$, $-\infty < v < \infty$ and $-\infty < r < \infty$. 
In the full Rindler spacetime, shown in figure~\ref{fig:diagram1}, these co-ordinates 
cover region~I, where $r>0$, region~III, where $r<0$, and their joint boundary, 
the right-going future branch of the Rindler horizon, where $r=0$. 

The analogue of the BMS-type supertranslations for the Rindler horizon 
in the metric \eqref{metricRindler} are 
given by the supertranslation vector
\begin{equation}
\Xi^a = \frac{1}{\kappa}[f(x,y), 0, -r \partial^A f(x,y) ]
\end{equation} 
which preserves the Bondi-type gauge 
\eqref{metricRindler} in the sense that $g_{rr} = 0$ and $g_{vr} =2$, 
and it also preserves the structure of the Rindler horizon, 
in the sense that ${\cal L}_{\Xi}g_{vv} = 0 + {\cal O}(r)$ and 
${\cal L}_{\Xi}g_{Av} = 0 + {\cal O}(r)$ \cite{Donn, Eling, Cai}. 
Joining the metric in 
Eq.\ \eqref{metricRindler} 
to the supertranslated metric along a shock wave propagating at $v=v_0$, we find that the perturbations to the Rindler metric $h_{ab} = {\cal L}_{\Xi}g_{ab}$ due to the shock wave are then given by
\begin{eqnarray}
{\cal L}_{\Xi}g_{Av} &=& h(v - v_0) 2r  \partial_A f  \\
{\cal L}_{\Xi}g_{AB} &=& h(v - v_0) 2 \left(\frac{r}{\kappa}\right) \partial_A \partial_B f\label{rindpert}
\end{eqnarray}
where again $h(v-v_0)$ is the 
Heaviside 
step function. 
Hence, the full metric describing the shock wave at $v=v_0$ passing through the Rindler horizon at $r=0$ is 
\begin{eqnarray}
ds^2 &=& - 2\kappa r dv^2 + 2 dvdr + 4r h(v-v_0) \partial_A f dvdx^A \nonumber \\
&& + \bigg( \delta_{AB} + 2 h(v-v_0) \frac{r}{\kappa} \partial_A \partial_B f \bigg) dx^A dx^B 
\label{eq:rindlershock1}
\end{eqnarray}  
Working to linear order in $f$, we find the linearised
stress-energy tensor of the shock wave to be
\begin{eqnarray}
T_{vv} &=& -r h^\prime \partial_A \partial^A f - \frac{r}{\kappa} h^{\prime \prime} \partial_A \partial^A f \nonumber \\
T_{vr} &=& \frac{h^\prime}{\kappa}  \partial_A \partial^A f \nonumber \\
T_{vA} &=& - h^\prime \partial_A  f \nonumber \\
T_{AB} &=& 2 \frac{h^\prime}{\kappa}  \partial_A \partial_B f  \;\;\;\;\;\;\;\;\; A=B \nonumber \\
T_{AB} &=& -2 \frac{h^\prime}{\kappa}  \partial_A \partial_B f  \;\;\;\;\;\; A\neq B
\label{rindlertab}
\end{eqnarray}
where the primed indices denote differentiation with respect to $v$. 
Note that $T_{ab}$ is by construction covariantly conserved.

Similarly to the Schwarzschild case, one could demand the surface stress-energy tensor to have an additional ${\bar \mu} h^\prime(v - v_0)$ term, with ${\bar \mu} >0$ a constant, in the null-null component $T_{vv}$ such that
\begin{eqnarray}
T_{vv} &=& {\bar \mu} h^\prime -r h^\prime \partial_A \partial^A f - \frac{r}{\kappa} h^{\prime \prime} \partial_A \partial^A f 
\label{rindlertabnew}
\end{eqnarray}
with all the other components as in 
Eq.\ \eqref{rindlertab}. 
The physical interpretation of ${\bar \mu}$ is the
surface energy density of the shock wave.  The corresponding metric perturbations 
in Eq.\ \eqref{rindpert} then get modified to 
\begin{eqnarray}
h_{Av} &=& h(v - v_0) 2r  \partial_A f  \\
h_{AB} &=& \frac{{\bar \mu}}{\kappa}\delta_{AB} h(v-v_0)\left( e^{\kappa(v-v_0)} - 1 \right) + h(v - v_0) 2 \left(\frac{r}{\kappa}\right) \partial_A \partial_B f
\end{eqnarray}

We emphasise that the linearised stress-energy tensor \eqref{rindlertab} 
has come by reverse-engineering a matter source for the 
Rindler supertranslation, 
and for generic $f$ this stress-energy tensor may  
not satisfy any of the usual energy conditions. 
However, when the surface energy term \eqref{rindlertabnew} is included, 
the stress-energy tensor can be made to satisfy the null energy condition for ${\bar \mu}>0$. 
This is all similar to what happens with the shock wave \eqref{schwsupermetric} in Schwarzschild, 
as is seen by comparing \eqref{eq:schw-stressenergy} 
to \eqref{rindlertab} and~\eqref{rindlertabnew}. 

We also emphasise a significant difference from the Schwarzschild case. 
In Schwarzschild the surface energy density term $\mu/4\pi r^2$ 
leads to a perturbation in the $g_{vv}$ metric component, but in Rindler 
it is the \emph{transverse\/} part of the metric that 
gets perturbed due to the surface density~${\bar \mu}$. 
We show in section \ref{memorysection}
that this leads to a drastic difference 
in the effect on the uniformly linearly accelerated trajectories: 
the effect in Rindler is a trajectory-dependent Lorentz boost, 
while the effect in Schwarzschild is an instability that knocks the trajectory 
away from stationarity and, for $\mu>0$, makes it fall into the black hole.

\section{Letaw-Frenet equations in curved spacetime} \label{linearsection}

To describe the memory effect for uniformly linearly accelerated observers, we need a covariant definition 
of such observers in a spacetime that is not necessarily flat. Below we motivate the need for such a construction.

In Minkowski spacetime, a 
uniformly accelerated trajectory may be defined as a timelike orbit of a Killing vector. 
These orbits were classified in terms of Lorentz-signature Frenet equations by Letaw \cite{Letaw}, 
and summaries in terms of the geometry of the Killing vectors are given in \cite{Sriram,Jorma}. 
As each of the trajectories is an orbit of a one-parameter isometry group, 
the magnitude of the proper acceleration is constant along the trajectory. 
The uniformly accelerated trajectories are hence a special 
class among worldlines on which the proper acceleration four-vector has constant magnitude. 
Note that if just the magnitude of the proper acceleration 
were fixed to a constant $\alpha>0$, 
the direction of the proper acceleration four-vector 
would still remain freely 
specifiable on the $S^2$ of radius $\alpha$ in the hyperplane orthogonal 
to the four-velocity, at each point on the worldline.

Among the uniformly accelerated trajectories, the orbits of a boost 
are called uniformly \emph{linearly} accelerated. 
In quantum field theory, a uniformly linearly accelerated observer
reacts to the Minkowski vacuum as if it were a thermal state \cite{Unruh}; by contrast, 
the observer's response under the other types of uniform acceleration 
is not expected to be thermal \cite{Letaw,Sriram,Jorma}. 
This special property of the boost can be attributed to 
the specific form of the entanglement between the field modes in the causally disconnected 
quadrants separated by the boost Killing horizon, 
and to the fact that each trajectory stays in one of the quadrants.\footnote{If the observer 
is direction-specific, 
thermality does however arise also for the accelerated trajectory constructed from a 
boost and a commuting spatial translation, but with an anisotropic temperature 
that contains a direction-dependent Doppler shift factor \cite{Townsend, San}.} 
In this paper we hence focus on observers of uniform \textit{linear} acceleration.

To address uniformly linearly accelerated observers in the presence of matter shock waves, 
we do however need to generalise the notion of uniform linear acceleration to a spacetime 
that is not necessarily flat. We now proceed to do this.

Letaw \cite{Letaw} showed that the construction of the generalised 
Frenet equations in Minkowski spacetime can be utilised to define analogues of the scalar curvature, 
the torsion scalar and the hypertorsion scalar for world lines in flat spacetime. 
In particular, the scalar curvature is the magnitude of the proper acceleration. 
The case of uniform linear acceleration then arises when the scalar curvature is fixed to a constant positive value while the torsion and hypertorsion scalars are taken to vanish.

To generalise the Letaw-Frenet construction to curved spacetime, we begin as in \cite{Letaw}
by defining four unit vectors forming an orthogonal tetrad using the Gram-Schmidt 
orthogonalisation procedure. These are defined at each point along the trajectory $x^a(\tau)$ of interest as
\begin{eqnarray}
V_0^a &=& u^a = \frac{dx^a}{d\tau} \nonumber \\
V_1^a &=& 
\frac{a^a}{|a|} \nonumber \\
V_2^a &=& \frac{|a|^2  w^a - |a|^2  \left( w^b u_b \right) u^a - \left( w^b a_b \right) a^a }{N} 
\nonumber \\
V_3^a &=& \frac{-1}{\sqrt{6}}\frac{\epsilon^{abcd}}{ \sqrt{-g}} V_{0b}V_{1c}V_{2d}
\label{Vdef}
\end{eqnarray}
where $a^a = u^b \nabla_b u^a $ , $w^a = u^b \nabla_b a^a$ and
$N = |a| \left( |a|^2 w_a w^a -  (a_a w^a)^2  + |a|^4 \right)^{1/2}$. 
Assuming $a^a \neq 0$ and $N \neq 0$, the four unit vectors of the tetrad by definition satisfy the following condition at the tangent space
at each event along the trajectory
\begin{equation}
V_{\alpha a} V^{ a}_{\beta} = \eta_{\alpha \beta} 
\label{Vproperty}
\end{equation} 
where the Greek indices label the respective unit vector. The generalised Letaw-Frenet equations 
then are
\begin{eqnarray}
u^b \nabla_b V_{\alpha}^a = K^\beta_\alpha V_\beta ^a
\label{Gfrenet}
\end{eqnarray}
where 
\begin{equation}
  K_{\alpha \beta}=
  \begin{pmatrix}
    0 & -{\cal K}(\tau) & 0 & 0 \\
    {\cal K}(\tau) & 0 & -{\cal T}(\tau) & 0 \\
    0 & {\cal T}(\tau) & 0 & -{\cal V}(\tau) \\
    0 & 0 & {\cal V}(\tau) & 0
  \end{pmatrix}
  \label{Kdefine}
\end{equation}
To arrive at \eqref{Kdefine}, 
we may proceed as in flat spacetime \cite{Letaw}, 
by taking the derivative of the orthogonality condition
\eqref{Vproperty} along $u^a$, using 
\eqref{Gfrenet} to deduce antisymmetry of $K_{\alpha \beta}$, 
and finally using \eqref{Vdef} to deduce that the $\alpha^{th}$ 
row in $K_{\alpha \beta}$ can have non-zero entries only in the columns with $\beta \leq \alpha + 1$.
The scalar quantities ${\cal K}(\tau)$, ${\cal T}(\tau)$ and ${\cal V}(\tau) $ can be straightforwardly identified as the analogues of the curvature scalar, torsion and the hypertorsion scalars respectively by simply constructing a local inertial frame around any event on the trajectory and matching the covariant scalars with those in the construction of Letaw's Frenet equations in flat spacetime. 
Note that ${\cal K}(\tau)$ is the magnitude of the proper acceleration, 
${\cal K}(\tau) =  (u^b \nabla_b V_{0a}) V_1^a = |a|$.

We may now define the curved spacetime analogue of uniform acceleration by requiring 
${\cal K}(\tau)$, ${\cal T}(\tau)$ and ${\cal V}(\tau)$ to be independent of $\tau$. 
Uniform \emph{linear} acceleration is defined as the special case in which ${\cal K}$ 
is strictly positive and 
${\cal T}$ and ${\cal V}$ vanish. 
For uniform linear acceleration, the only non-trivial Frenet equation is then the 
equation of motion for the normalised acceleration vector,  
\begin{eqnarray}
u^b \nabla_b V_{1}^a &=& K^0_1 V_0 ^a \nonumber \\
\Rightarrow \; \; u^b \nabla_b a^a = w^a &=& |a|^2 u ^a 
\label{linearFrenet}
\end{eqnarray}
The above equation was also obtained in \cite{Rindler} by generalising the differential- geometric characteristics of a rectangular hyperbola in Minkowski spacetime to curved spacetimes.
As a technical caveat, we should note that the tetrad \eqref{Vdef} 
is not well defined for uniform linear acceleration because 
the formula for $V_2^a$ takes the ambiguous form $0/0$, using \eqref{linearFrenet}.  
This can be remedied by defining  
a binormal $V_{2,3}^{ab}$ to the plane of $V_{0a}$ and $V_{1a}$ by
\begin{eqnarray}
V_{2,3}^{ab} = \frac{-1}{\sqrt{6}}\frac{\epsilon^{abcd}}{ \sqrt{-g}} V_{0c}V_{1d}
\label{binormal}
\end{eqnarray}
In the space spanned by this binormal, one can choose two unit vectors such that the orthonormality condition in Eq.\ \eqref{Vproperty} still holds. The analysis then proceeds as above and once again the only non-trivial Frenet equation is Eq.\ \eqref{linearFrenet}. 
The consistency of the setup can be verified by considering the change in the binormal $V_{2,3}^{ab}$ along the trajectory,
\begin{eqnarray}
u^e \nabla_e V_{2,3}^{ab} &=& \frac{-1}{\sqrt{6}}\frac{\epsilon^{abcd}}{ \sqrt{-g}} \left( u^e \nabla_e V_{0c}\right) V_{1d} + \frac{-1}{\sqrt{6}}\frac{\epsilon^{abcd}}{ \sqrt{-g}}V_{0c} \left( u^e \nabla_e V_{1d}\right) \nonumber \\
&=& 0
\end{eqnarray}
where the first term vanishes since $u^e \nabla_e V_{0c}$ is parallel to $V_{1d}$ and the second term vanishes on using the constraint equation Eq.\ \eqref{linearFrenet}.  The vanishing of the $u^e\nabla_e V_{2,3}^{ab}$ confirms that Eq.\ \eqref{linearFrenet} is consistent with the vanishing of the torsion and hypertorsion scalars as required. One can note that the explicit form of the unit vectors $V_2^a$ and $V_3^a$ in this case are not needed.

A heuristic way to arrive at the the constraint equation 
Eq.\ \eqref{linearFrenet} is as follows. 
In order to impose uniform linear acceleration, we 
demand that the acceleration vector $a^a$ has a constant positive magnitude, 
and any change in $a^a$ lies in the plane spanned by $u^a$ and $a^a$. 
The latter condition implies 
\begin{equation}
u^b\nabla_b a^a = w^a = p_1 u^a +  p_2 a^a
\label{pedaplane}
\end{equation} 
where $p_1 = - u_a w^a$ and $p_2 = a_a w^a/|a|^2$, using $u_a a^a=0$. 
As $|a|$ is constant, we have 
\begin{align}
0 = u^b \nabla_b (a^a a_a) 
= 2a_a w^a = 2 p_2 |a|^2
\end{align}
which implies $p_2=0$. 
As $u_a a^a=0$, we have  
\begin{align}
0 = u^b \nabla_b (u^a a_a)  = |a|^2 + u_a w^a 
\end{align} 
which implies $p_1 = |a|^2$. Collecting, we obtain the constraint 
\begin{equation}
w^a  - |a|^2 u^a = 0
\label{linearcon}
\end{equation}
which is identical to 
Eq.\ \eqref{linearFrenet}. 
Using 
Eqs.\ \eqref{pedaplane} and \eqref{linearcon} and the constancy of $|a|$, we further see that all further of derivatives of $u^a$ 
lie in the plane spanned by $u^a$ and $a^a$.

\section{The memory effect for accelerated observers}\label{memorysection}

In this section, we investigate the gravitational memory effect of supertranslations on
uniformly linearly
accelerated trajectories. We consider 
in turn the Rindler and Schwarzschild spacetimes with 
supertranslational hair implanted by an asymmetric shock wave as discussed in section \ref{hairsection}. 
Starting with a family of uniformly linearly accelerated 
trajectories that follow the orbits of a single Killing vector 
before the shock wave, the task is to find what these 
trajectories have become after the wave has passed.
We begin with the Rindler spacetime. 

\subsection{Rindler spacetime}
We work in the Bondi-type gauge where the Rindler metric before the shock wave, 
$v<v_0$  can be expressed as in \eqref{metricRindler}, 
\begin{equation}
ds^2 = - 2\kappa r dv^2 + 2 dvdr  + \delta_{AB}  dx^A dx^B
\label{eq:rindlerbondi2}
\end{equation} 
The boost Killing vector in these co-ordinates is $\bar{\xi}^a = \kappa^{-1}(1,0,0,0)$. 
Without loss of generality, let us consider a representative Rindler trajectory for $v<v_0$ with 
the world-line $x^a(\tau) = [\tau/\sqrt{2\kappa r_c}, r_c, x^A_c]$ where $\tau$ is the proper time along the trajectory and $r_c$, $x^A_c$ 
are the initial values.
It is easy to check that 
the trajectory is linearly uniformly accelerated 
in the sense of Eq.\ \eqref{linearFrenet}, 
and $|a| = \kappa/\sqrt{2\kappa r_c}$.

Let an asymmetric shock wave at $v=v_0$ with the stress-energy tensor \eqref{rindlertab} impinge on the Rindler horizon. Working to linear order in the perturbation, we can investigate separately the memory effect due to the surface energy density term ${\bar \mu} h(v-v_0)$ and the memory effect due to the supertranslation perturbation terms. We first consider the case without the surface energy density term and set ${\bar \mu} = 0$. The resultant metric is as in \eqref{eq:rindlershock1}, 
\begin{eqnarray}
ds^2 &=& - 2\kappa r dv^2 + 2 dvdr + 4r h(v-v_0) \partial_A f dvdx^A \nonumber \\
&& + \bigg( \delta_{AB} + 2 h(v-v_0) \frac{r}{\kappa} \partial_A \partial_B f \bigg) dx^A dx^B 
\label{metric2}
\end{eqnarray}    
with the corresponding supertranslation vector
\begin{equation}
\Xi^a = \frac{1}{\kappa}[f(x,y), 0, -r \partial^A f(x,y) ]
\label{eq:Xi-Rindler}
\end{equation}

We assume that $h(v-v_0) = \lambda {\cal H}(v-v_0)$ where $\lambda$ is a small 
dimensionless perturbative parameter and ${\cal H}(v-v_0)$ 
is the Heaviside step function. To determine the trajectory on 
and after the shock wave for $v \ge v_0$, we make 
for the trajectory's four-velocity the ansatz
\begin{equation}
u^a = 
\left[ \frac{1}{\sqrt{2\kappa r}}, 0, 
{\cal E} (v) \frac{\sqrt{2\kappa r}}{\kappa} \partial^A f 
\right]
\label{velocitysol}
\end{equation}
where ${\cal E} (v)$ is of first oder in $\lambda$ and 
must be determined from \eqref{linearFrenet}. 
The acceleration vector is
\begin{equation}
a_a = \left[ 0, \frac{\kappa}{2\kappa r}, 
\frac{1}{\kappa}\frac{d h}{dv}\partial_A f  + \frac{1}{\kappa}\frac{d {\cal E}}{dv}\partial_A f 
\right] 
\end{equation}
and it satisfies 
\begin{eqnarray}
a^2  = \frac{\kappa^2}{2\kappa r} + {\cal O}(\lambda^2)
\end{eqnarray}
This shows that the ansatz \eqref{velocitysol} 
is consistent with keeping the magnitude the acceleration constant to linear order. 
What remains is to impose the constraint \eqref{linearFrenet}, 
which to linear order takes the form 
\begin{equation}
0 = w_a  - a^2 u_a = 
\left [ 
0, 0, \frac{1}{\kappa \sqrt{2 \kappa r} }\frac{d^2 h}{dv^2}\partial_A f  
+ \frac{1}{\kappa \sqrt{2 \kappa r} }\left(\frac{d^2 {\cal E}}{dv^2} 
- \kappa^2 {\cal E}\right)\partial_A f 
\right] 
\end{equation}
The equation for ${\cal E}(v)$ is hence 
\begin{eqnarray}
\frac{d^2 {\cal E}}{dv^2} - \kappa^2 {\cal E}  = - \frac{d^2 h}{dv^2}
\end{eqnarray}
and matching to the orbits of the boost Killing vector 
$\bar{\xi}^a$ before the wave gives the initial condition 
${\cal E}(v) = 0$ for $v<v_0$. 
The solution is 
\begin{eqnarray}
{\cal E}(v)  =  - h(v-v_0) \cosh[\kappa (v-v_0)] 
\end{eqnarray}
The four-velocity vector of the trajectory is hence 
\begin{equation}
u^a = 
\left[
\frac{1}{\sqrt{2\kappa r}}, 0, - h(v-v_0) \cosh[\kappa (v-v_0)]  
\frac{\sqrt{2\kappa r}}{\kappa} \partial^A f 
\right]
\label{velocitysolrindler}
\end{equation}
Integrating \eqref{velocitysolrindler} to first order in $\lambda$, 
we find that the trajectories are 
\begin{align}
x^a(\tau) =  
\left[
\frac{\tau}{\sqrt{2\kappa r_c}},r_c, 
x^A_c - h \! \left(\frac{\kappa (\tau-\tau_0)}{\sqrt{2\kappa r_c}}\right)
\partial^A f(x^A_c) 2r_c \sinh\left(\frac{\kappa (\tau-\tau_0)}{\sqrt{2\kappa r_c}}\right) \right]
\label{trajectoryrindler}
\end{align}
where $\tau_0$ is the proper time at $v=v_0$. 

For $v<v_0$, the trajectories \eqref{trajectoryrindler} are by construction 
integral curves of the boost Killing vector $\bar{\xi}^a = \kappa^{-1} (1,0,0,0)$. 
What are these trajectories for $v>v_0$? 

For $v>v_0$, working to linear order in $\lambda$, 
the perturbed metric is related to the Rindler metric by a diffeomorphism generated 
by the supertranslation vector $\Xi^a$ \eqref{eq:Xi-Rindler}. As 
${\cal L}_{\Xi}\bar{\xi}^a  = 0$, it follows that 
$\bar{\xi}^a$ is a boost Killing vector also for $v>v_0$. 
Assume now that $f$ is generic. 
It is then immediate from \eqref{velocitysolrindler}
that the trajectories \eqref{trajectoryrindler} 
are \emph{not\/} orbits of $\bar{\xi}^a$. 
Further, we have verified that a vector field
parallel to the velocity field \eqref{velocitysolrindler}, 
of the form 
\begin{equation}
q^a = Q(r,x^A)\left [1, 0,  {\cal E} (v) 2r \partial^A f \right]
\ , 
\end{equation}
satisfies Killing's equation to linear order in $\lambda$ only when 
${\cal E}(v) = 0$ and $Q(r,x^A)$ is a constant. 
This means that
the velocity field \eqref{velocitysolrindler} 
is not parallel to a Killing vector, and the 
trajectories \eqref{trajectoryrindler}  
do not constitute a family of integral curves of a Killing vector field. 

However, the Letaw-Frenet construction at $v>v_0$ guarantees that each trajectory in the family 
\eqref{trajectoryrindler} is the orbit of \emph{some\/} boost Killing vector. 
This means that the Killing vector must differ from trajectory to trajectory. 
When the velocity vector in 
\eqref{velocitysolrindler} is transformed to a set of standard 
Minkowski co-ordinates $(T,X,Y^A)$, it takes the form 
\begin{eqnarray}
U^a &=& \bigg[\cosh\left(\frac{\kappa \tau}{\sqrt{2 \kappa r_c}} - \frac{\log[\kappa r_c]}{2}\right), \sinh\left(\frac{\kappa \tau}{\sqrt{2 \kappa r_c}} - \frac{\log[\kappa r_c]}{2}\right), \nonumber \\
&& \hspace{2ex}
h \! \left(\frac{\kappa (\tau-\tau_0)}{\sqrt{2\kappa r_c}}\right) \alpha^A \cosh\left(\frac{\kappa \tau}{\sqrt{2 \kappa r_c}} -\kappa v_0 \right)  \bigg] 
\label{velocitymin}
\end{eqnarray}
where $\alpha^A = \frac{\sqrt{2\kappa r_c}}{\kappa} \partial^A f(x^A_c)$, 
and we have used \eqref{trajectoryrindler} to express the 
velocity vector in terms of the proper time~$\tau$.
From \eqref{velocitymin} we see that 
a trajectory with given $r_c, x^A_c$ is an integral curve of a boost 
Killing vector that is obtained by applying to $\bar{\xi}^a$ the Lorentz boost 
\begin{equation}
  \Lambda^a{}_b=
  \begin{pmatrix}
    1 & 0  & \alpha^Y \cosh\beta & \alpha^Z \cosh\beta \\
    0 & 1 & -\alpha^Y \sinh\beta & -\alpha^Z \sinh\beta \\
    \alpha^Y \cosh\beta & \alpha^Y \sinh\beta & 1 & 0 \\
    \alpha^Z \cosh\beta & \alpha^Z \sinh\beta & 0 & 1
  \end{pmatrix}
  \label{lorentzboost}
\end{equation}
where $\beta= (1/2)\log[\kappa r_c] - \kappa v_0$. 
Note that the magnitude and direction of the boost \eqref{lorentzboost} 
depend on $r_c$ and $x^A_c$. 

Collecting, we have shown that implanting a supertranslational hair on the 
Rindler horizon by our matter shock wave boosts a family of Rindler 
trajectories in a way that differs from trajectory to trajectory, 
and this trajectory-dependence carries 
a memory of the planar inhomogeneity of the wave. 
This is the gravitational memory effect for 
uniformly linearly accelerated observers.

To end this subsection, we return to the case of positive ${\bar \mu}$. 
Working to linear order, 
we can set the perturbations due to the supertranslational terms to zero. 
The relevant metric in this case is 
\begin{eqnarray}
ds^2 &=& - 2\kappa r dv^2 + 2 dvdr  \nonumber \\
&& + \bigg( \delta_{AB} + \frac{{\bar \mu}}{\kappa}\delta_{AB} h(v-v_0)\left( e^{\kappa(v-v_0)} - 1 \right) \bigg) dx^A dx^B 
\label{umetric}
\end{eqnarray}    
We now find that the velocity vector field 
\begin{equation}
u^a = \left[ 
\frac{1}{\sqrt{2\kappa r}}, 0, 0,0
\right]
\label{eq:u-mubar}
\end{equation}
has $|a|= \kappa/\sqrt{2 \kappa r_c}$ and 
satisfies the Letaw-Frenet constraint \eqref{linearFrenet} for all~$v$. 
For $v < v_0$, the trajectories are orbits of the boost 
Killing vector $\bar{\xi}^a$. 
For $v > v_0$, the metric \eqref{umetric} is flat to linear order, and 
when the velocity vector \eqref{eq:u-mubar} is transformed to a 
standard set of Minkowski co-ordinates $(T,X,Y^A)$, it takes the form 
\begin{eqnarray}
U^a &=& \bigg[\frac{X}{\sqrt{X^2-T^2}},\  \frac{T}{\sqrt{X^2-T^2}}, \ \frac{\bar \mu e^{-\kappa v_0}x_c^A (T+X)}{2\sqrt{X^2-T^2}}\bigg] 
\label{boost3}
\end{eqnarray}
where $X > |T|$. 
We see again that after the wave has passed, each trajectory is an integral curve of a boost, 
but the boost differs from trajectory to trajectory. The effect can be interpreted 
as a focusing due to the energy density in the wave.

\subsection{Schwarzschild spacetime}\label{schwsection}

We now turn to the infalling linearised
shock wave in the supertranslated Schwarzschild spacetime, 
and to its consequences for a family of trajectories that are static before the wave 
and are continued to the future of the wave as uniformly linearly accelerated trajectories.
We proceed as in Rindler. Working in the Bondi gauge where the supertranslational hair implanting shock wave is infalling in the Schwarzschild black hole metric, the complete metric reads 
as in \eqref{schwsupermetric},
\begin{eqnarray}
ds^2 &=& - \left(1 - \frac{2M}{r} - h(v - v_0)\frac{2\mu}{r}  -  h(v - v_0)\frac{MD^2 C}{r^2}\right) dv^2 + 2dv dr \nonumber \\
&& -  h(v - v_0) D_A \left( 2C - \frac{4MC}{r} + D^2C \right) dv d\Theta^A \nonumber  \\
&& + \left( r^2\gamma_{AB} + h(v - v_0) 2rD_A D_B C -  h(v - v_0)r\gamma_{AB} D^2C \right) d\Theta^A \Theta^B \nonumber \\
\end{eqnarray}
For notational simplicity, we write $V = 1 - 2M/r$. 

We consider trajectories that are static for $v<v_0$, 
$x^a(\tau) = [\tau/\sqrt{V(r_c)}, r_c, \theta^A_c]$, where $\tau$ is the proper time along the trajectory and $r_c$, $\Theta^A_c$ are the initial co-ordinates. These trajectories are uniformly linearly accelerated in the sense of section \ref{linearsection}, and the magnitude of the acceleration is 
$|a| = V^\prime(r_c)/2\sqrt{V(r_c)}$. 
We continue the trajectories across and to the future of the shock wave by keeping them uniformly linearly accelerated. 

We first consider the effect from the shock wave mass term $\mu$, taking the perturbed metric to be
\begin{eqnarray}
ds^2 &=& - \left(1 - \frac{2M}{r} - h(v - v_0)\frac{2\mu}{r} \right) dv^2 + 2dv dr +  r^2\gamma_{AB} d\Theta^A \Theta^B \nonumber \\
\end{eqnarray}
and assuming $\mu>0$. As the overall effect of $\mu$ is to increase the mass of the black hole, 
we may anticipate the initially static trajectories to become unstable on crossing the 
shock wave and to fall into the black hole. We now verify that this is the case within the perturbative treatment. 
A nonperturbative treatment could be given by the methods of \cite{Barrabes:1991ng,poisson-toolkit}. 

Working to linear order in the perturbation, we assume that $h(v-v_0) = \lambda {\cal H}(v-v_0)$ 
where $\lambda$ is a small  dimensionless perturbative parameter and ${\cal H}(v-v_0)$ is the 
Heaviside
step function. 
To find the trajectory for $v \ge v_0$, we assume $r-2M \gg 2\mu$ and seek the velocity vector field by the ansatz
\begin{equation}
u^a = \left[\frac{1+ \frac{h(v-v_0) \mu}{rV} + {\cal E} (v)}{\sqrt{V}}, {\cal E} (v), 0,0 \right]
\label{velocitysol2}
\end{equation}
where ${\cal E} (v)$ is to be determined. 
For the magnitude of the acceleration vector, we find   
\begin{eqnarray}
|a|^2 =  \frac{V^{\prime 2}}{4V}  + \left( \frac{V^{\prime}}{V}\right) {\cal E}^\prime +   \left( \frac{V^{\prime}}{rV^2}\right) \mu h^\prime +   \left( \frac{V^{\prime}}{r^2 V}\right) \mu h +   \left( \frac{V^{\prime 2}}{2r V^2}\right) \mu h 
\label{schwacc}
\end{eqnarray}
where the prime denotes differentiation with respect to the argument, that is, 
$V^\prime = dV/dr$, $h^\prime = dh/dv$ and ${\cal E}^\prime = dh/dv$. 
The constraint \eqref{linearFrenet} gives 
\begin{eqnarray}
0 = w_a  - a^2 u_a &=& \frac{1}{4V^{5/2}r^2}[0, 4r\mu h^{\prime \prime} + 2 \mu r h^{\prime} V^\prime 
+ 4r^2 V {\cal E}^{\prime \prime}  \nonumber \\
&& + 2 {\cal E}r^2 V^2 V^{\prime \prime} + 4 \mu V h^{\prime} - r^2 V V^{\prime 2} {\cal E} ,0,0]
\label{eq:Schw-linearfrenet}
\end{eqnarray}
Differentiation of \eqref{schwacc} with respect to $v$ shows that 
\eqref{eq:Schw-linearfrenet} implies constancy of $|a|$. 
The only equation that needs to be solved is hence \eqref{eq:Schw-linearfrenet}. 

Writing ${\cal E}(v)$ in terms of $(1/\sqrt{V})dr/dv$, \eqref{eq:Schw-linearfrenet} gives 
\begin{equation}
\frac{d^2r_\epsilon}{dv^2} -V(r_c)\left( |a|^2 -  \frac{V^{\prime \prime}(r_c)}{2} \right) r_\epsilon 
= \frac{- \mu h^\prime}{r_c} - \frac{\mu V(r_c) h}{r^2_c} - \frac{\mu V^\prime(r_c) h}{2 r_c} 
\end{equation}
where $r_\epsilon = r - r_c$. 
With the initial condition $r_\epsilon(v) =0$ for $v < v_0$, the solution is 
\begin{eqnarray}
r &=& r_c - h(v-v_0) \frac{\mu }{r_c \beta} \sinh \left( \beta (v-v_0)\right) \nonumber \\
&& - h(v-v_0) \left( \frac{\mu V}{r_c^2} +  \frac{\mu V^\prime}{2 r_c} \right) \frac{2}{\beta^2} \sinh^2\left( \frac{\beta}{2}(v-v_0) \right)
\label{schwtraj}
\end{eqnarray}
with $\beta^2  = V(r_c)\left( |a|^2 -  V^{\prime \prime}(r_c)/2 \right)$. 
The solution has exponential runaway and will eventually exit the
regime in which the linearised treatment is valid, but the signs in \eqref{schwtraj}
show that the trajectory will start to fall towards the black hole, as we anticipated. 

When the supertranslation terms in the metric are added, the Rindler analysis suggests that the trajectories
will carry a memory of the spherical anisotropy of the wave, 
and a generic trajectory will either fall into the black 
hole or escape to infinity.

\section{Discussion}\label{discsection}

In this paper we have demonstrated and quantified a gravitational memory effect due to a matter shock wave that 
implants supertranslational hair on a Rindler horizon. 
We considered a family of observers who follow the integral curves of a Lorentz boost prior to the wave, 
and we assumed the observers to continue as uniformly linearly accelerated across the wave, 
in the sense of a curved spacetime generalisation of the 
Letaw-Frenet uniform linear acceleration in flat spacetime \cite{Letaw}. 
After the wave has passed, we find that 
each observer still follows the orbit of a boost Killing vector, but 
this boost differs from trajectory to trajectory, and the trajectory-dependence carries 
a memory of the planar inhomogeneity of the wave. We also considered a matter shock wave that 
implants supertranslational hair 
on the Schwarzschild spacetime \cite{HPS2}, 
showing that a similar memory effect on initially static uniformly 
linearly accelerated trajectories exists but involves an instability that makes 
the trajectories fall into the black hole or escape to the infinity. 

In Schwarzschild, the linearised stress-energy tensor of the 
supertranslation-implementing 
shock wave involves a Dirac delta on a null hypersurface \cite{HPS2}. 
In Rindler, by contrast, we found that the linearised stress-energy tensor 
of the supertranslation-implementing shock wave, in addition to a Dirac delta term,  also involves a \emph{derivative\/} 
of the Dirac delta on a null hypersurface. 
Studying the shock wave beyond the linearised theory \cite{Barrabes:1991ng,poisson-toolkit,Geroch:1987qn}
could hence be significantly more challenging in Rindler than in Schwarzschild. 

While our discussion was classical, 
it is motivated by the potential of supertranslations 
as a solution to the black hole information paradox \cite{HPS1}. 
As the classical memory effect due to Rindler supertranslations
involves a trajectory-dependent boost, the Killing horizons of the uniformly linearly accelerated 
trajectories in the future of the shock wave are boosted with respect to each other. 
In terms of spacetime regions separated by the Rindler horizons, 
some of the degrees of freedom that prior to the shock wave were 
inaccessible to a particular Rindler observer become hence accessible in the future of the shock wave, 
and vice versa. This leads us to anticipate that the classical memory effect due to the 
Rindler supertranslations has a counterpart in the thermal aspects of 
Rindler space quantum field theory, and we plan to address this effect 
in a future paper \cite{Sanjorma}.

\section*{Acknowledgments}

SK thanks the University of Nottingham for hospitality and
the Department of Science and Technology, India, for partial
financial support. 
JL was supported in part by the 
Science and Technology Facilities Council (Theory Consolidated Grant ST/J000388/1).

\end{document}